\documentstyle[11pt,aaspp]{article}
\newcommand{\BE}{\begin{equation}}
\newcommand{\BEAL}{\begin{eqnarray}}
\newcommand{\EE}{\end{equation}}
\newcommand{\EEAL}{\end{eqnarray}}
\def\kms{kms\&{-1}}
\def\ls{L\_{\odot}}
\def\ms{M\_{\odot}}
\def\ao{a_{o}}
\def\aoet{$\ao=2~10^{-8}~cms\&{-2}$}

\def\_#1{_{\scriptscriptstyle #1}}
\def\&#1{^{\scriptscriptstyle #1}}

\def\vr{v\_{r}}

\def\s{\sigma}
\begin{document}
\title{MOND and the seven dwarfs}
\author{ Mordehai Milgrom}
\affil{Department of Condensed-Matter
 Physics, Weizmann Institute of Science  76100 Rehovot, Israel}
\begin{abstract}
 \cite{g94} had recently analyzed the data on seven
dwarf spheroidals, and concluded that these disagree with the predictions
 of MOND. We contend that this conclusion is anything but correct.
With new data for three of the dwarfs the observations are in compelling
 agreement with the predictions of MOND. Gerhard found MOND $M/L$ values
 that fall around a few solar units, as expected if  MOND is a valid
alternative to dark matter--while the Newtonian $M/L$ values are in the
 range of about 10 to 100 solar units. Gerhard's sole cause for complaint
was that some of his MOND $M/L$ values were still outside the range of
 ``reasonable'' stellar values, say between 1 and 3 (two were too
 high--about 10--and one too low). This, we say, was {\it easily}
 attributable to uncertainties in the data, such as in the velocity
 dispersions, luminosities, core radii, etc., and in the assumptions that
 underlie the analysis, such as isotropic velocity distribution,
 light-distribution fits, no tidal effects, etc.. There are now new,
 much-improved
 determinations of the velocity dispersions for three of the dwarfs,
 in particular for the two dwarfs for which
Gerhard found high MOND $M/L$ values: The preliminary, high value
 for Leo II has been superseded by a much smaller
value (by a factor of two) that gives a MOND $M/L$ value of about
1.5 (instead of about 9). There is also a new determination of the
velocity dispersion of Ursa Minor which brings down the MOND $M/L$ value
considerably. The one small MOND $M/L$ value that
Gerhard found is also shown to be a red herring. In fact, within just
the quoted errors on the velocity dispersions and the luminosities,
the MOND $M/L$ values for all seven dwarfs are perfectly consistent with
 stellar values, with no need for dark matter.

\end{abstract}
\keywords{Dwarf spheroidals, dark matter, modified dynamics}
\section{Introduction}
\par
Low-surface-density galaxies provide crucial test cases for the modified
dynamics--MOND (Milgrom 1983a,b). Low surface density
is tantamount to low accelerations, and it is the basic tenet of MOND
that low-acceleration systems must evince large mass discrepancies
(or in the common way of thinking, large amounts of dark matter).
Relevant systems include very diverse types, such as dwarf spirals,
dwarf irregulars, low-surface-brightness (LSB) galaxies of normal
or large size,  and the dwarf-spheroidal (DS) neighbors of the Milky way.
\par
Long before relevant data was available for any of these classes,
\cite{mil83b} predicted that such galaxies will show large mass
discrepancies; and that, unlike higher-surface-density galaxies, the
discrepancy will appear starting from the
centre of the galaxy, not
only beyond some transition radius, as the accelerations in such
systems are small already at the centre.
These predictions have been amply born out by recent observations of
dwarf spirals [\cite{csv88,mb88,cb89,jc90,lsv90,pcb90,bbs91}], of
 dwarf irregulars [\cite{lsy93,mil94}], and of LSB galaxies. For a sample
of six of these last (the ilk of Malin 1) the results of \cite{vdh93}
 imply $M/L$ values roughly between 10 and 75 solar units, at the last
 measured points on the rotation curves (with large uncertainties);
  MOND analysis gives values consistent with values of a few solar
 units (Milgrom, unpublished).
\par
About dwarf spheroidals--which are the subject of this paper--we find in
 \cite{mil83b}:
``...we predict that when velocity-dispersion data is available for the
(spheroidal)
dwarfs, a large mass discrepancy will result when the conventional
dynamics is used to determine the masses. The dynamically determined
mass is predicted to be larger by a factor of order 10 or more than that
which is accounted for by stars.'' This prediction was made solely on the
basis of the low surface brightness of the dwarfs.
\par
We know now that, indeed, the Newtonian $M/L$ values
   deduced for DS are, typically, between 10 and 100 solar units.
 [For recent compilations and further references to earlier work see
 \cite{mat94}, and \cite{vmok95}.] These are still very uncertain due to
 both observational and procedural uncertainties
 [see e.g. \cite{l90,pk90,pry94}].
Indeed, the very assumption of virial equilibrium, which underlies
the mass determination, has been questioned.
\par
If the dynamical mass in the dwarfs is dominated by stars, then,
if we determine the masses of the dwarfs using MOND we should get
$M/L$ values that are typical of their stellar populations.
\par
Applying MOND analysis to seven DS (Draco, Carina, Ursa Minor, Sextans,
Sculptor, Fornax, and Leo II) \cite{g94}
claims that the predictions of MOND for these dwarfs are in conflict with
 the observations. His analysis extends and updates
an analysis for five of the dwarfs by \cite{gs92}, who reach
qualitatively similar, unfavorable conclusions.

\par
Quoth \cite{g94}: `` Two results are immediately apparent: Objects like
 Fornax require very low $M/L$ in MOND, while others like UMi
 still require
significant dark matter. Between the dwarf spheroidals, $M/L$ must vary
by a factor of at least 20 even in MOND.''
\par
In fact, ``objects like Fornax'' stands for Fornax alone, for which
Gerhard found central
 $M/L$ values between 0.2 and 0.3; and ``objects like UMi''
stands for UMi ($M/L\sim 10-13$) and Leo II ($M/L\sim 9$).
 For Carina, Sextans, and
Sculptor Gerhard found, by the appropriate MOND estimator,
 $M/L$ values between 1 and 3, and for Draco he found $M/L\sim 6$.
\par
Even with these results, Gerhard's adverse conclusion was unwarranted,
as he has been well aware of the large uncertainties still
besetting the analysis.
 We discuss these in more detail below, and show that within the quoted
uncertainties available to \cite{g94} his MOND M/L values were consistent
with stellar values (he does not discuss the error range).
\par
In vindication of the above statement
there exist now new, high-quality determinations of the
velocity dispersions that improve dramatically the agreement with MOND:
(i) The determination of the dispersion for
 Leo II by \cite{vmok95} that supersedes the
preliminary value [\cite{mat94}] used by \cite{g94} gives a MOND $M/L$
value of about 1.5, instead of the 9 that Gerhard obtained.
(ii) A new determination of
the dispersion in UMi [\cite{hgic94b}] that is much smaller than that
used by Gerhard, and which gives UMi a stellar MOND $M/L$ value.
(iii) An updated value of $\s$ for Draco would have given Gerhard an
 $M/L$ value of 4 instead of 6.
\par
All in all, the agreement with MOND is now as good as could be expected
in light of the remaining uncertainties.

In \S 2 we briefly describe the various MOND $M/L$ estimators, and
 discuss the uncertainties involved in their determination.
 In \S 3 we consider individual dwarfs with their idiosyncrasies.

\section{MOND $M/L$ estimators and their uncertainties}

\par
\cite{gs92}, and \cite{g94} consider four types of MOND $M/L$ estimators;
 two of them concern central values, and two global values--the
 difference being analogous to that in the Newtonian case.
 Within each  pair, one estimator is based on the assumption
that the dwarf is isolated and unaffected by the external field of the
Milky Way; we refer to this as the global value.
The other estimator is based on the oppositely extreme
 assumption that
the dwarf is dominated by the external field:
the acceleration with which the system falls
in the external field is large compared with internal accelerations.
In the latter case, the internal dynamics in MOND is very nearly
Newtonian, with an effective gravitational constant that is determined
by the external acceleration [\cite{mil86}]--we
 call this limit, after \cite{gs92},
the quasi-Newtonian case.
 All accelerations relevant to the
dwarf dynamics are much smaller than the acceleration constant $\ao$.
In this case the effective gravitational constant in the quasi-Newtonian
limit is simply $G\ao/a\_{ex}$, where $a\_{ex}$ is the acceleration of
the system in the external field.
Thus, in this case, the MOND $M/L$ value is obtained by multiplying the
Newtonian value by $a\_{ex}/\ao$.
 We can write $a\_{ex}=V^2/R$, where
$R$ is the galactocentric distance of the dwarf, and $V$ is the
rotational velocity at the position of the dwarf. As did \cite{gs92},
and \cite{g94},
we take $V=V\_{\infty}=220~\kms$ for all the dwarfs.
\par
A measure of the degree of isolation in the above sense is provided
by the parameter
 \BE \eta\equiv {3\s^2/2r_c\over V^2\_{\infty}/R}
\approx{a\_{in}\over a\_{ex}}, \label{i} \EE
where $r_c$ is the core radius, and $\s$ is the (mean) line-of-sight
 velocity dispersion.
\par
Some of the dwarfs are borderline, with $\eta\sim 1$. In this case
the two estimators (quasi-Newtonian, and isolated) should give similar
$M/L$ values. Neither limit is valid in this case, but the correction
to $M/L$ due to this fact is not large (it works to increase $M/L$
 somewhat, by at most about 40 percent).
\par
 For the isolated case \cite{gs92} use an estimator of the central, MOND
$M/L$ value of the form $M/L\propto \s\_{0}\&{4}/\ao I(0)r_c^2$,
where $\s\_{0}$ is the central velocity dispersion, and
 $I(0)$ is the central surface brightness.
 For the global estimator one uses the MOND mass-velocity relation
 \BE M={9\over 4}{\s\_{3}^4\over G\ao}, \label{ii} \EE
where $\s\_{3}$ is the {\it three-dimensional} rms velocity dispersion
for the whole system.
 \cite{mil94} derived this relation for an arbitrary, low-acceleration
 ($a\_{in}\ll\ao$), stationary, isolated system in the formulation of
MOND given by \cite{bm84};
it had been derived earlier for spherical systems by \cite{gs92}
 [and earlier yet for isothermal spheres in \cite{mil84}].
\par
 If the system is globally isotropic, i.e. has the same line-of-sight
 rms dispersion, $\s$, in all directions, we have $\s\_{3}=3\&{1/2}\s$
and can write
 \BE M={81\over 4}{\s^4\over G\ao}; \label{iii} \EE
this is the relation used all along for the global isolated case.
\par
Regarding uncertainties, MOND estimators are, naturally,
 free of the uncertainly related to the unknown distribution of the dark
 matter relative to the visible one. In fact, in the quasi-Newtonian case
the assumption  that ``mass follows light'' is appropriate, as the MOND
effect here is predominantly
 to increase the value of the gravitational constant. The quasi-Newtonian
estimators are, otherwise, subject to the same uncertainties that beset
the Newtonian analysis, and which have been much emphasized in the
literature [e.g. \cite{l90,pk90,cal92,pry94,vmok95}].
\par
Among the observational parameters of the dwarfs, the
photometric (structural)
 parameters--central surface brightness, core radius, tidal radius,
total luminosity--are notoriously difficult to determine.
\cite{cal92}--whence come much of the photometric parameters used
by \cite{g94}--discuss these problems in detail.
The velocity dispersions had also been rather uncertain, but there
seems to be rapid and  significant progress in this regard
with many more stellar velocities measured, and increased
control of contribution from binaries. The isolated MOND estimators
are more sensitive to uncertainties in the dispersions as they are
proportional to the fourth power of the latter.
The central MOND estimators (isolated or quasi-Newtonian)
are very sensitive to assumptions on the unknown distribution of orbit
orientations in the system in analogy to the Newtonian case [see e.g.
\cite{l90,pk90}].
 Most of the estimates mentioned below assume an isotropic
distribution.
The global, isolated MOND estimator does not require knowledge of the
orbit distribution, but makes use of the rms dispersion for the whole
system, when many times only central values are available. In only
two cases (Fornax and Leo II) are there, at the moment, some assurances
that the quoted value can be used as the rms dispersion.
The global luminosity value is, on the other hand, less certain than
the central luminosity (say within the half brightness radius), which
enter the determinations of central values.
 \par
The possible presence of tidal-disruption effects on the velocity
dispersions has also been raised: At least some of the
dwarfs may be undergoing tidal disruption to some degree, a reckoning
without which fact would lead to an overestimate of the mass.
\cite{g94}, and
\cite{vmok95} summarize the pros and cons of this actually being an
important effect, and give references to earlier discussions.
The most cogent counter argument is that no systematic gradients of
velocities are found as would be expected if tidal disruption is
responsible for much of the velocity dispersion.
\cite{mop93} define a parameter $X$  that measures the potential
susceptibility of the dwarf to tidal effects (ratio of the central
 density in the dwarf to the average galactic density within the
galactocentric distance of the dwarf).
 In the analysis we ignore such effects.

\section{ MOND $M/L$ values of individual dwarfs}
\par
Here we estimate the MOND $M/L$ values of the individual dwarfs. We use
all along \aoet, as did \cite{gs92} and \cite{g94},
 to facilitate comparison (all
derived $M/L$ values scale like $\ao\&{-1}$). Unless stated otherwise
 we take all dwarf parameters as used by \cite{g94} [who, in turn,
 takes them from \cite{mat94}].
\vskip 0.2truecm
\centerline{\it Sculptor}
 This dwarf has $\eta\approx 0.5$,
 which points to domination by the
external field. For this situation Gerhard found a central,
 MOND $M/L=1.2$, based on the Newtonian value $(M/L)\_{N}=12$
 [\cite{mat94}].
 The range resulting from errors in $\s$ alone--
 $\s=6.8\pm0.8~\kms$--is $0.9\le M/L \le1.5$. The central surface
 brightness comes originally from \cite{cal92}, who give an error
 estimate of $\pm0.3$ mag/arcsec$^2$,
 increasing the range to $0.7\le M/L \le2.0$.

\vskip 0.2truecm
\centerline{\it Sextans}

With the quoted dispersion, $\s=6.2\pm0.8$
 this has an $\eta\approx 0.3$, and is thus most probably
dominated by the external field. It is also among those with the poorest
isolation value by the $X$ criterion. For this case \cite{g94} found a
 central, MOND $M/L$ of 1.5 [based on a Newtonian value of 18 from
\cite{mat94}, who, in turn, quotes \cite{mop93}].
 The range corresponding to the error
 in $\s$ is $1.1\le M/L \le1.9$.
The value of the central surface brightness used by Gerhard
originates with \cite{cal92} who had estimated the error to be $\pm0.5$
mag/arcsec$^2$, further broadening the range to $0.7\le M/L \le3.0$.
\par
Note that there are significant differences between the structural
 parameter values used by different authors.
 We, with \cite{g94}, use those given in
\cite{mat94,mop93}; but compare \cite{smtow93}, and \cite{hgic94a}.

\vskip 0.2truecm
\centerline{\it Carina}

Here $\eta\approx 0.6$. For the external-field-dominated case, which
 seems more appropriate, Gerhard found a MOND, central value $M/L=3.6$
(1.8 for the isolated case). This corresponds to the central value of
$\s=6.8\pm 1.6$  [\cite{mop93,mat94}]. The range corresponding to the
error on $\s$ alone is $2.1\le M/L\le 5.5$. The quoted relative error on
the central luminosity surface density is $\pm 0.3$ mag/arcsec$^2$
 [\cite{mop93}],
further broadening this range to $1.6\le M/L\le 7.2$.
The Newtonian, V-band, central $M/L$ value from which these are derived
 [\cite{mop93}] is $(M/L)\_{N}=39\pm 23$.

\vskip 0.2truecm
\centerline{\it Draco}

With the value of the dispersion used by \cite{g94},
 $\s=10.2\pm 1.8~\kms$,
 this  dwarf has $\eta\approx 1.3$ and it is plausible to treat
it as isolated, for which case Gerhard gets a central MOND
value of $M/L=6.1$ (5.8 for the quasi-Newtonian case). The range that
corresponds to the quoted measurement errors on $\s$
 alone is $2.8\le M/L\le 11.7$. [\cite{gs92} had
 obtained for this dwarf MOND
$M/L$ values of 17 and 10 instead of the 6.1 and 5.8 that Gerhard
 found--highlighting the state of flux the analysis had been in.]
\par
 Recently, \cite{poa95} have obtained
a somewhat smaller value for the dispersion, with a much smaller error,
 based on measurements of many
more stellar velocities: $\s=9.2\pm 0.8$, which gives an isolated
 MOND $M/L=4$,
with a range $2.8\le M/L\le 5.6$, corresponding to the error on $\s$
 alone (I could not find an error quoted for the central surface
 brightness).

\vskip 0.2truecm
\centerline{\it Leo II}

 \cite{g94} used the preliminary value of $\s=13.8~\kms$
 given in \cite{mat94}, to obtain an isolated, central, MOND $M/L$ of
 9.3. However, the quoted error on $\s$ was $\pm5.5~\kms$
 allowing an $M/L$ value as small as 1.2.
\par
At any rate, there is now a new determination of the dispersion
of Leo II that definitely supersedes the above mentioned.
\cite{vmok95} find $\s=6.7\pm1.1~\kms$; they give this as the rms
for Leo II, not the central value. Dividing the data between three
radial bins \cite{vmok95} find no significant change of $\s$ with radius.
 We thus use it to derive the more
sound, global $M/L$ value using relation (\ref{iii}).

 This dwarf has $\eta=1.4$ (with the updated $\s$),
 and is the most isolated in the MOND
sense (with respect to the external acceleration).
 It is also, by far,  the least likely to be affected by tidal
 effects.
We treat it as isolated, remembering that the effect of the external
 field, which may be present,
 would be to increase somewhat the MOND $M/L$ value.
\par
The MOND mass for Leo II for the above value of $\s$
is $M=(1.5\&{+1.25}\_{-0.77})10^6~\ms$.
 With the (V-band) luminosity given
in \cite{vmok95}, $L=(9.9\pm 0.2)10^5~\ls$, we get a global
$M/L=1.5$ with an error range of between 0.7 and 3.3.
 The global, Newtonian $M/L$ found by \cite{vmok95}
is $(M/L)\_{N}=11.1\pm3.8$.

\vskip 0.2truecm
\centerline{\it Ursa Minor}

 For this dwarf \cite{g94} uses the value
 $\s=12.0\pm 2.4~\kms $ given in \cite{mat94}. However, in a more recent
paper, \cite{hgic94b} find a much lower value of
 $\s=7.5\&{+1.0}\_{-0.9} \kms $, based on measurements of many more
velocities. This smaller value is consistent with the recent result of
\cite{poa95} who find
 $\s=8.9\pm 0.8~\kms $. With the high value of $\s$ we get
 $\eta\approx 1$, which makes UMi a borderline case. Gerhard found,
central, MOND $M/L$ values of 10 for the isolated case and 13 for the
quasi-Newtonian case. The ranges corresponding to the quoted errors on
$\s$ alone are
 $4.1\le M/L\le 20.7$, and  $8.3\le M/L\le 18.7$, respectively.
The value of the central surface brightness used by Gerhard
originates with \cite{cal92} who had estimated the error to be $\pm0.5$
mag/arcsec$^2$, further broadening the error range on $M/L$.
Gerhard himself, in his discussion of Newtonian $M/L$ values for the
dwarfs says that the range of values for UMi is about 20--100; yet, for
the derivation of his MOND value he used a Newtonian value of 110, at
the uppermost end of this range.
\par
With the smaller
value of $\s$ we obtain $\eta\approx 0.3$, and it is
reasonable to treat UMi as dominated by the external field. In this case
we get, updating only $\s$, and leaving the other parameter values as
 used by Gerhard, $M/L=5.1$, instead of 13. The range corresponding to
 the quoted errors on $\s$ is
 $3.9\le M/L\le 6.5$. The central, Newtonian value \cite{hgic94b} find is
 $(M/L)\_{N}=59\&{+41}\_{-25}$.
 The errors on the central surface brightness,
core radius, and total luminosity broadens the above given ranges
(to say nothing of the effects of wrong assumptions such as
isotropic orbits). For example, the error \cite{hgic94b} quote for
the surface luminosity density is about $\pm50\%$, which reduces
the lower limit on $M/L$ to 2.6.
 Furthermore, \cite{hgic94b} find some contribution from rigid rotation
to the velocity dispersion--confirmed by \cite{poa95}--which reduces
 the value of the pure dispersion to  $\s=6.7\&{+0.9}\_{-0.8}~\kms$
 further  reducing the above estimates of the MOND $M/L$ values.

\vskip 0.2truecm
\centerline{\it Fornax}

 Here $\eta\approx 0.8$, and so Fornax is a borderline case
as regards MOND isolation. It is deemed relatively immune to tidal
 effects by the $X$ criterion. Fornax is also unique among the seven in
 that
line of sight dispersions have been measured to relatively large radii.
Preliminary analysis of Mateo and Olszewski (private communication)
shows that $\s$ remains constant down to the tidal radius. Thus, we can
safely use the MOND global mass-velocity relation eq.(\ref{iii})
 that is
less uncertain than the relation employing central quantities.
Assuming an isolated Fornax we get from eq.(\ref{iii}), and the quoted
$\s=11\pm2~\kms$, a MOND mass
of $M=1.1~10^7~\ms$, with a range corresponding to the quoted error on
$\s$ of $(0.5-2.2)~10^7~\ms$. We find in the literature
 rather different values quoted for
the (V-band) luminosity of Fornax. \cite{g94} uses $L\_{V}=2.5~10^7\ms$
 which originates with
 \cite{cal92} who estimate the error to be $\pm .5$ mag/arcsec$^2$;
 \cite{gs92} use a value three times smaller of $L\_{V}=8~10^6\ms$
quoted from \cite{pry92}. With the former value of $L$ the range of
MOND $M/L$ values is 0.2--0.9 with only the error on $\s$, and $0.1-1.4$
with the error on $L$ included.
With the latter value of $L$ the range is 0.6--2.7.

 \cite{g94} does not give global values, but
 \cite{gs92} do, and they find, like us, values of order
1 (with a somewhat smaller dispersion of $10~\kms$): 0.95 for the
isolated case, and 0.74 for the purely quasi-Newtonian case.

 \cite{g94} found small MOND $M/L$ values (0.2-0.3), based on a Newtonian
value of about 5.
Some idea of the uncertainties can be gotten from the range
of acceptable Newtonian values \cite{mow91} cite:
 $5.3\le (M/L)_{N}\le 26$,
 stating that this large range is dominated by uncertainties
in the structural parameters of Fornax. No improvement of the situation
has since been made in this regard. The low MOND $M/L$ values Gerhard
 found are based on a Newtonian value at the very low end of this range.
\par
\cite{mow91} argue that Fornax must
 have a stellar $M/L$ value ``about half
that observed in globular clusters or in the range 0.5--1.5'', because
it is characterized by a relatively young population, which is expected
to have rather smaller $M/L$ values. The MOND $M/L$ values we find
and even the lower values Gerhard found are, when considering the
uncertainties, definitely consistent with such stellar values.

\par
In summary of this section,
we find that with the updated values of the velocity dispersions, and
 within just the quoted measurement errors on the velocity dispersions
 and luminosities, the observations of all seven dwarfs are
perfectly consistent with MOND stellar
 $M/L$ in the range of 1 to 3 solar units. The kinematics of the dwarfs
can thus be understood with MOND dynamics and with no need for
 dark matter.
\acknowledgements
I Thank Mario Mateo and Ed Olszewski
 for providing results prior to publication.

\end{document}